\documentclass[prb,twocolumn,showpacs]{revtex4}
\usepackage{graphicx}
\usepackage{amssymb}
\usepackage{amsmath}

\begin{document}

\title{Solution of the Holstein polaron anisotropy problem}

\author{Andreas Alvermann}
\author{Holger Fehske}
\affiliation{
Institut f{\"u}r Physik, Ernst-Moritz-Arndt-Universit{\"a}t Greifswald,
17489 Greifswald, Germany }
\author{Stuart A. Trugman}
\affiliation{Theoretical Division, Los Alamos National Laboratory, Los
Alamos, New Mexico 87545, USA}

\begin{abstract}
We study Holstein polarons in three-dimensional anisotropic materials.
Using a variational exact diagonalization technique we provide highly accurate
results for the polaron mass and polaron radius.
With these data we discuss the differences between polaron formation
in dimension one and three, and at small and large phonon frequency.
Varying the anisotropy we demonstrate how a polaron evolves from a
one-dimensional to a three-dimensional quasiparticle.
We thereby resolve the issue of polaron stability in
quasi-one-dimensional substances and clarify to 
what extent such
polarons can be described as one-dimensional objects.
We finally show that even the local Holstein interaction
leads to an enhancement of anisotropy in charge carrier motion.
\end{abstract}


\maketitle

\section{Introduction}
In materials with a local coupling of the
charge carriers to optical phonons 
a tendency towards formation of polarons prevails.
How polarons form, and which properties they possess, has been
intensively studied using the Holstein model
(for a recent review see Ref.~\onlinecite{FT07}).
Much work was devoted to the question how polaronic properties depend
on dimensionality.
For zero phonon frequency polaron formation
is known to be fundamentally different in dimension one (1D) and three
(3D):
While in 1D a polaron exists for all coupling strengths,
in 3D a transition from a free electron to a self-trapped polaron
occurs at a finite coupling strength~\cite{KM93}. 
While the 3D polaron is small, in 1D large polarons with a size of many lattice constants may exist~\cite{WF98}.
This qualitative difference is peculiar to the static case.
At finite phonon frequencies no strict self-trapping transition occurs~\cite{GL91}.
Instead a smooth crossover from a light electron to a heavy small polaron takes place with increasing coupling strength.
Only in the adiabatic regime of small phonon frequencies,
large and heavy polarons can be found in 1D~\cite{Adiab}.
In the anti-adiabatic limit of large phonon frequencies polaron
properties do not depend on dimension.
For intermediate phonon frequencies numerical studies~\cite{KTB02} reveal that
in 3D polarons 
form as smaller and heavier quasiparticles
if compared to the 1D case, but a clear distinction as for zero phonon frequency or in the adiabatic regime is not found.

Most work on the Holstein polaron problem was performed for
1D or isotropic systems.
Much less attention is paid to the question how polaron properties in
anisotropic materials, e.g. molecular crystals~\cite{Ger06}, interpolate between the 1D and 3D behaviour.
In particular, even polarons in quasi-1D systems, such as in
organic conductors like the Bechgaard salts,
are frequently described by the 1D Holstein model,
neglecting the possibility of electron motion transverse to a
quasi-1D chain.
For zero phonon frequency Emin showed~\cite{Emi86} that
already tiny electronic transfer integrals perpendicular to the chain
suffice to destabilize a large 1D polaron, resulting in a free
electron as in 3D below the self-trapping transition.
This result not only puts the existence of large (adiabatic) polarons in real materials in doubt,
but also raises the question to which degree polarons in quasi-1D
systems can be described as true 1D polarons.
In the present work we study to which extent the scenario described for zero phonon frequency translates to finite phonon frequencies beyond the extreme adiabatic regime.

\section{Anisotropic Polaron Model}

To address this fundamental question, we consider the 3D Holstein
polaron model with anisotropic hopping 
\begin{equation}\label{Ham}
\begin{split}
  H = & - \sum_{i=x,y,z} t_i \sum_\mathbf{n} \,  \left( c^\dagger_{\mathbf{n}+
    \mathbf{e}_i} c_\mathbf{n} + c^\dagger_\mathbf{n} c_{\mathbf{n}+
    \mathbf{e}_i} \right)
 \\
  & - \sqrt{\varepsilon_p \omega_0} \sum_\mathbf{n} (b^\dagger_\mathbf{n} +
  b_\mathbf{n}) c^\dagger_\mathbf{n} c_\mathbf{n} + \omega_0
  \sum_\mathbf{n} b^\dagger_\mathbf{n} b_\mathbf{n} \;.
\end{split}
\end{equation}
Here $\mathbf{n}$ labels the sites of a cubic lattice,
and $\mathbf{e}_i$ for $i=x,y,z$ denotes the unit
lattice vector in the respective direction.
$\omega_0$ is the phonon frequency, and $\varepsilon_p$ the
electron-phonon coupling strength.
For the electron transfer integrals, we set $t_x=t_\parallel$,
$t_y=t_z=t_\perp$.
As $t_\perp/t_\parallel$ grows,
the system described by the Hamiltonian~\eqref{Ham} evolves from a 1D
($t_\perp/t_\parallel=0$) to an isotropic 3D system ($t_\perp/t_\parallel=1$).
For intermediate values of $t_\perp/t_\parallel$, the system is
anisotropic but symmetric with respect to the $y,z$-direction, i.e.,
the chains in a quasi-1D system ($t_\perp/t_\parallel \ll 1$) are
oriented along the $x$-direction.

In addition to Ref.~\onlinecite{Emi86},
the anisotropic Holstein model~\eqref{Ham} has been studied in
Refs.~\onlinecite{RBL99b,LI07} in certain limiting cases, or using
approximate variational techniques.
At present, no conclusive answers for small-to-intermediate phonon frequency have been given.
In this work, we present accurate results for the infinite
system using the variational exact diagonalization technique developed in Ref.~\onlinecite{BTB99}.
The error of all results is generally much less than $1\%$, 
and about $1\%$ for the polaron radius in the crossover regime at small $\omega_0$ in 3D.
Our presentation concentrates on small-to-intermediate phonon frequencies $\omega_0/t_\parallel=0.1 \dots 2.0$. We justify this choice in the next paragraphs.

\section{Numerical Results and Discussion}
\subsection{1D and 3D cases}
The destabilization of a polaron at small $t_\perp/t_\parallel$ occurs only if the polaron is large.~\cite{Emi86}
We therefore begin our discussion with the basic question in what cases large polarons exist at finite phonon frequencies. 
Using the electron-phonon correlation function
$ \chi(\mathbf{r}) = \sum_\mathbf{n} \langle \psi_0|
(b^\dagger_{\mathbf{n}+\mathbf{r}}+b_{\mathbf{n}+\mathbf{r}})
c^\dagger_\mathbf{n} c_\mathbf{n} |\psi_0\rangle$,
the polaron radius, which measures the extension of the phonon cloud surrounding the electron, is defined as
\begin{equation}\label{polrad}
  R_\parallel = \left[ \frac{ \sum_\mathbf{r} (r_x)^2 \chi(\mathbf{r})
  }{2 \sum_\mathbf{r} \chi(\mathbf{r}) }   \right]^{1/2} \;,
\end{equation}
and similarly for $R_\perp$ with $r_x$ replaced by $r_y$ or $r_z$.
Note that for the Holstein model, $\sum_\mathbf{r} \chi(\mathbf{r}) = 2\sqrt{\varepsilon_p/\omega_0}$.

\begin{figure}
\includegraphics[width=\linewidth]{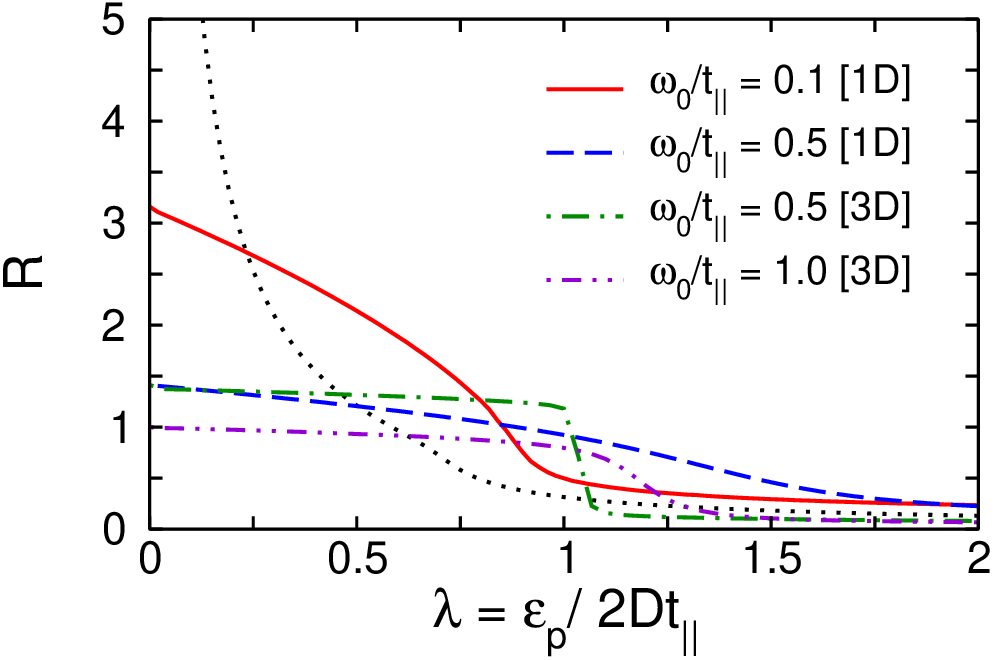}
\caption{(Color online) 
Polaron radius $R$ as a function of coupling $\lambda$, for dimension $D=1$ ($t_\perp=0$) and $D=3$ (isotropic case $t_\parallel=t_\perp$).
The dotted curve gives the radius for $\omega_0=0$ and 1D, which is defined according to Eq.~\eqref{polrad} using the electron density
$|\psi(\mathbf{r})|^2$ instead of the correlation
function $\chi(\mathbf{r})$.
}
\label{Fig1}
\end{figure}

How the distinctive features of polaron formation found for zero phonon frequency persist at small $\omega_0/t_\parallel$ is manifest in a comparison of the 1D ($t_\perp/t_\parallel=0$) and 3D ($t_\perp/t_\parallel=1$) curves for $R =R_\parallel=R_\perp$ in Fig.~\ref{Fig1},
given as a function of $\lambda = \epsilon_p / 2 D t_\parallel$.
The steep decrease of $R$ in the vicinity of $\lambda=1$ for 3D, $\omega_0/t=0.5$ is a precursor of the self-trapping transition for $\omega_0=0$,
which takes place at $\lambda^{3D}_c \approx 0.90$.
In 1D the polaron radius decreases steadily with $\lambda$, and no
crossover region can be identified.
Although the latter behaviour is reminiscent of the 1D large-to-small polaron crossover at $\omega_0=0$,
we observe that even for small phonon frequency $\omega_0/t_\parallel=0.1$ 
no large polaron is found.
In Ref.~\onlinecite{RBL99} it was shown that, independent of dimension, for weak coupling the polaron radius approaches
\begin{equation}
\lim_{\varepsilon_p \to 0} \; R_\parallel = \sqrt{t_\parallel/\omega_0} 
\end{equation}
(compare the 1D and 3D curve for $\omega_0/t_\parallel=0.5$ in Fig.~\ref{Fig1}).
The polaron radius is bounded by that limiting value, 
e.g., for $\omega_0 / t_\parallel = 0.1$, at most $R_\parallel < 3.2$.
To obtain a large polaron of radius greater than 5 or 10, 
$\omega_0/t_{\parallel}$ must be less than 0.01 to 0.04
for weak coupling $\epsilon_p/\omega_0, \lambda \ll 1$.
Furthermore the polaron shrinks rapidly with increasing coupling.
Even for $\omega_0=0$ a coupling of, say, $\lambda \gtrsim 0.25$ leads to a polaron radius of only $R \lesssim 2$.
The condition on the phonon frequency this becomes 
even more restrictive if we ask for large polarons with substantial mass,
in contrast to the light large polarons at weak coupling.

\begin{figure}
\includegraphics[width=\linewidth]{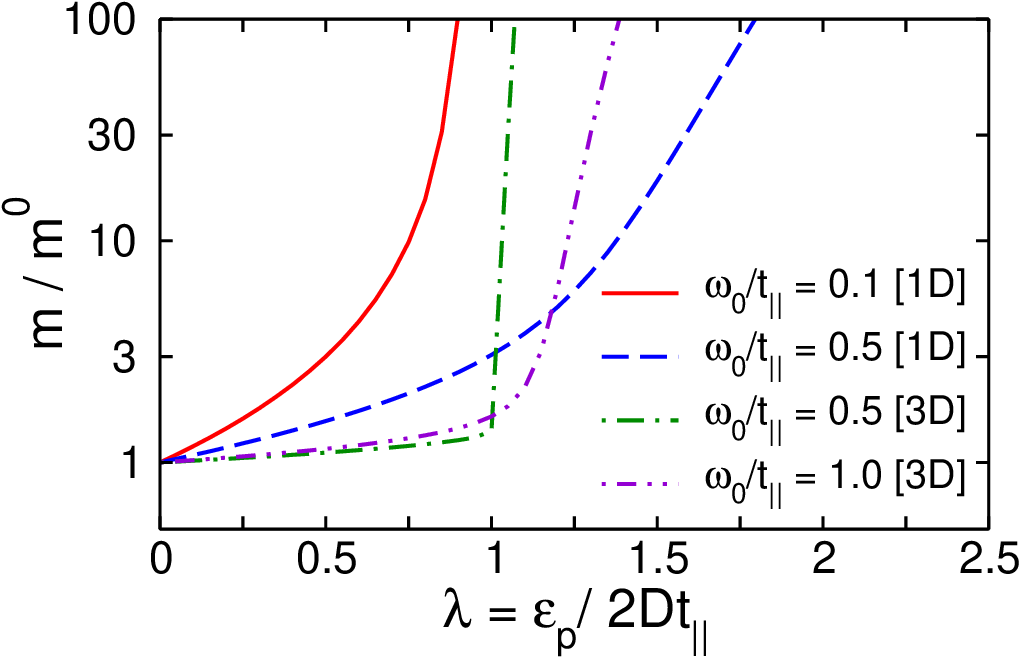}
\caption{(Color online) 
Polaron mass $m$ as a function of coupling $\lambda$, for 
the parameters as in the previous figure.
}
\label{Fig2}
\end{figure}

The polaron mass,
which is probably the most important quantity to characterize a polaron, 
is defined as
\begin{equation}
  m_i^{-1} = \frac{\partial^2 E(\mathbf{k})}{\partial k_i^2}
  \Big|_{\mathbf{k}=0} \;,
\end{equation}
where $E(\mathbf{k})$ is the groundstate energy 
at momentum
$\mathbf{k}$.
We set $m_\parallel = m_x$, and $m_\perp = m_y = m_z$.
In the absence of electron-phonon coupling ($\varepsilon_p=0$), 
the masses are given by $m^0_\parallel = 1/2t_\parallel$,
$m^0_\perp = 1/2t_\perp$.
At $\omega_0=0$ the (3D) transition
from a free electron to a self-trapped polaron coincides with a jump
of $m_i$ from $m_i=m_i^0$ (below the transition) to $m_i=\infty$ (above the
transition).
For finite $\omega_0$, $m$ depends analytically on the coupling
strength $\varepsilon_p$~\cite{GL91}.

In Fig.~\ref{Fig2} we show the polaron mass 
$m=m_\parallel$ as a function of the coupling strength
$\lambda=\varepsilon_p/2Dt$, for different $\omega_0$
in dimension $D=1,3$.
Remember that the data for $m$ 
correspond to the exact numerical solution of the Holstein model.
As for the radius the distinctive differences of polaron formation in 1D and 3D
are manifest in a comparison of the corresponding curves for $\omega_0/t_\parallel=0.5$.
While in 1D the polaron mass increases steadily with $\lambda$,
a steep increase of $m_\parallel$ occurs in the vicinity of $\lambda^{3D}_c$ for 3D.
One can therefore locate, for small phonon frequency
$\omega_0/t_\parallel \lesssim 1$ in 3D, the crossover from a light
particle to a heavy polaron in the region $\lambda \simeq 1$,
while in 1D no crossover region can be identified.
The differences between 1D and 3D vanish for larger phonon frequencies
$\omega_0/t_\parallel \gtrsim 1$ (not shown here), where the mass indicates
the continuous evolution from a light to a heavy polaron with
increasing $\lambda$.
In the anti-adiabatic regime $\omega_0/t_\parallel \gg 1$, 
the relation $m_i/m_i^0 = \exp(g^2)$ with $g^2=\varepsilon_p/\omega_0$ holds
independent of dimension.

The behaviour of the 1D polaron mass for small phonon frequency 
is reminiscent of the large-to-small polaron crossover in the 
$\omega_0=0$-limit.
However, for $\omega_0=0$ the mass of the 1D polaron is strictly infinite for any coupling strength in 1D. 
The $\omega_0=0$-result therefore does not provide a quantitative prediction
for the polaron mass at any finite phonon frequency.
Owing to that restriction, it is also not sufficient to establish the existence of large polarons with substantial mass renormalization even for the adiabatic case $\omega_0/t_\parallel \ll 1$.
Instead we see in Fig.~\ref{Fig2}, that even for a small phonon frequency $\omega_0/t_\parallel=0.1$ the mass of the 1D polaron is still close to unity for those coupling strengths $\lambda \lesssim 0.25$, for which a large polaron is found for sufficiently small $\omega_0/t_\parallel$. 
To obtain polarons with large mass and large radius therefore requires small coupling strengths $\lambda \ll 1$ (to allow for a large radius) and even smaller phonon frequencies $\omega_0 / t_\parallel \ll \lambda$ (to allow for a large mass).
Consequently, such heavy adiabatic large polarons may exist only at extremely small phonon frequencies $\omega_0 / t_\parallel \lesssim 0.01$.
We note that 
(i) in general no large-to-small
polaron crossover occurs even at small phonon frequencies in 1D,
(ii) 
the adiabatic regime of large (heavy) polarons occupies only the very narrow parameter region of tiny phonon frequencies,
thus
(iii) the destabilization of large quasi-1D polarons at small $t_\perp/t_\parallel$ Emin obtained for $\omega_0=0$ can be expected to be relevant only in that narrow region.
Qualitatively, we expect strong deviations from the findings for $\omega_0=0$
even for small-to-intermediate phonon frequencies beyond the extreme adiabatic regime.

\begin{figure}
\includegraphics[width=\linewidth]{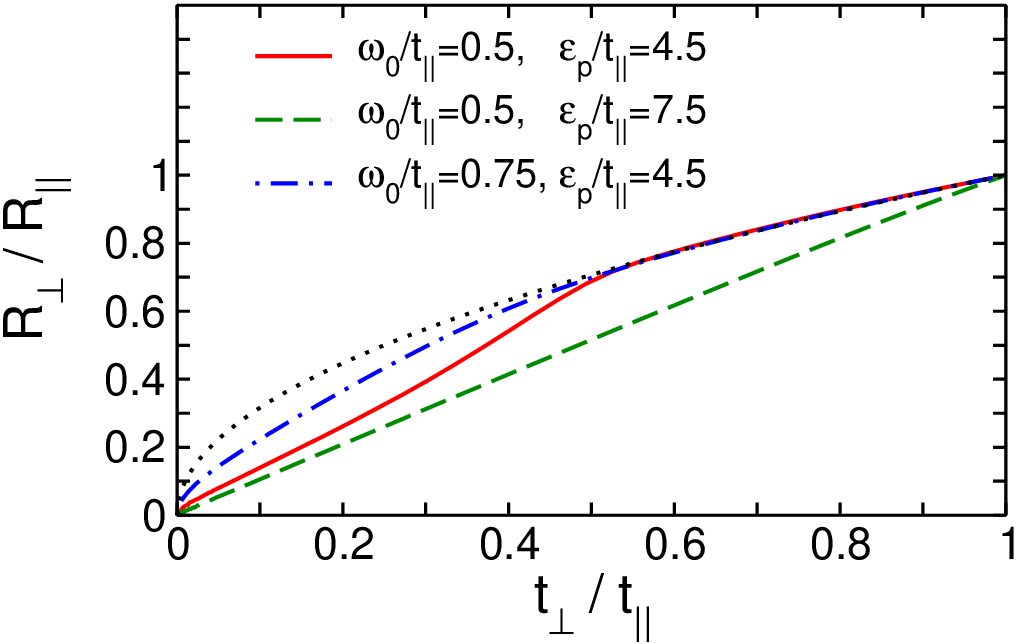}
\caption{(Color online) 
Ratio $R_\perp/R_\parallel$ of polaron radii
  as a function of anisotropy $t_\perp/t_\parallel$, for
  $\omega_0/t_\parallel$ and $\varepsilon_p/t_\parallel$ as indicated.
Here the dotted curve gives the weak-coupling result $R_\perp/R_\parallel=\sqrt{t_\perp/t_\parallel}$.}
\label{Fig3}
\end{figure}
\subsection{Spatial anisotropic case}
Let us now begin with the quantitative discussion of the anisotropic case.
Since in the extreme adiabatic regime of large (heavy) polarons
the question of polaron properties in anisotropic materials 
was already settled by Emin~\cite{Emi86} we exclude this narrow region from our further consideration.
Instead we concentrate on the broader regime of small-to-intermediate phonon frequencies.

Obviously, the polaron radius does not indicate per se to which degree the polaron ceases to be a true 1D particle for small $t_\perp/t_\parallel$.
We may instead use the ratio $R_\perp/ R_\parallel$ (see Fig.~\ref{Fig3}), which we interpret as a measure of the
``one-dimensionality'' of the polaron:
For $R_\perp/ R_\parallel=0$ ($R_\perp/ R_\parallel=1$), the polaron
is a fully 1D (isotropic 3D) object.
We see that already for small $t_\perp/t_\parallel$ the ratio $R_\perp/
R_\parallel$ deviates significantly from zero,
but $R_\perp/R_\parallel$ is a continuous function of $t_\perp/t_\parallel$ in contrast to the behaviour in the $\omega_0=0$-limit.
For weak coupling, we recover $R_\perp / R_\parallel =
\sqrt{t_\perp/t_\parallel}$, as can be derived by perturbation 
theory~\cite{RBL99}.
For strong coupling, the asymptotic behaviour
$R_\perp / R_\parallel = t_\perp/t_\parallel$ is approached.
For intermediate coupling ($\varepsilon_p/t_\parallel=4.5$ in the figure) we observe how the curve interpolates between the strong-coupling straight line at smaller $t_\perp/t_\parallel$ and the weak-coupling square root at larger $t_\perp/t_\parallel$.
To understand this observation,
we shall now use the polaron mass to further clarify how the polaron properties change with $t_\perp/t_\parallel$.

\begin{figure}
\centering
\includegraphics[width=0.8\linewidth]{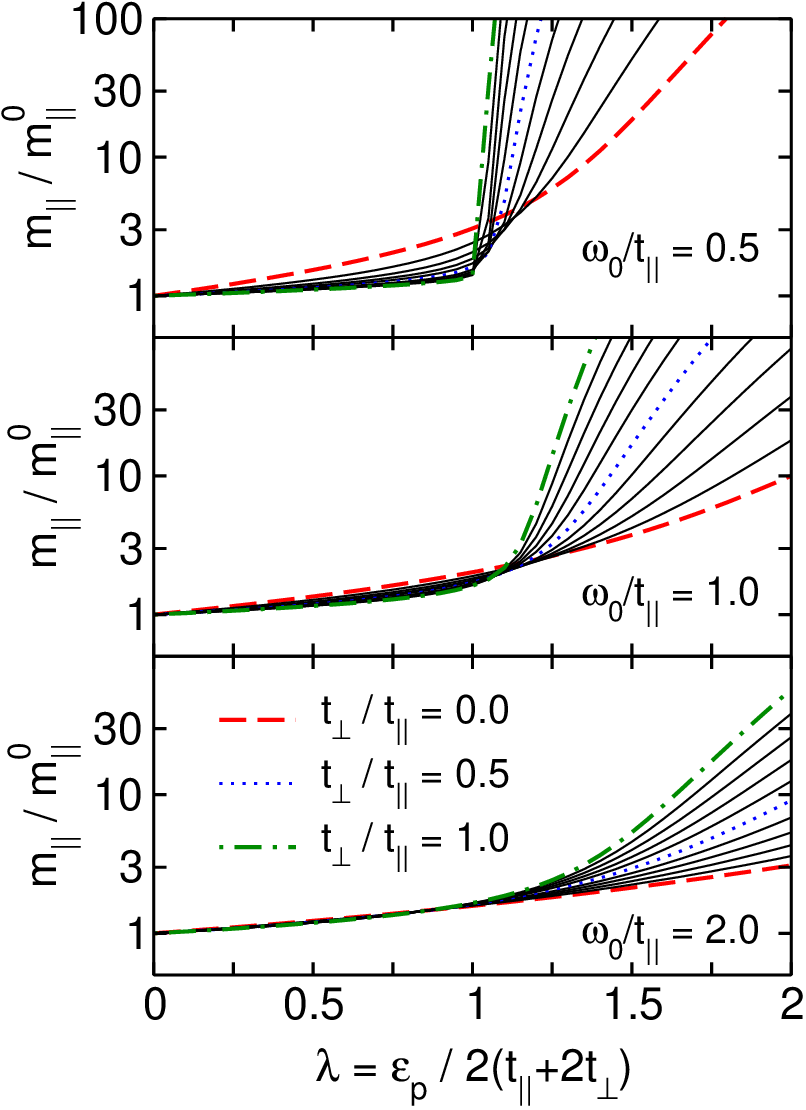}
\caption{(Color online) Polaron mass $m_\parallel$ for different anisotropy
  $t_\perp/t_\parallel$, and phonon frequencies $\omega_0$ as indicated.
In each panel, eleven curves for $t_\perp/t_\parallel= 0.0 \dots 1.0$
in steps of $0.1$ are shown.
Note that for fixed $\lambda=\varepsilon_p/2(t_\parallel+2t_\perp)$ the
coupling $\varepsilon_p$ depends on $t_\perp$.}
\label{Fig4}
\end{figure}

In Fig.~\ref{Fig4} we show the polaron mass 
$m_\parallel$ as a function of the coupling strength
$\lambda=\varepsilon_p/2(t_\parallel+2t_\perp)$, for different $\omega_0$
and anisotropy $t_\perp/t_\parallel$. 
We see that the curves for intermediate values of
$t_\perp/t_\parallel$ smoothly interpolate between the 1D and 3D case.
In contrast to the $\omega_0=0$-limit a sudden transition from 1D to 3D behaviour at small $t_\perp/t_\parallel$ is missing.
Nevertheless, for $\omega_0/t_\parallel=0.5$ and $\lambda>1$, the mass
$m_\parallel$ increases by a huge factor if $t_\perp/t_\parallel$
grows from zero to one.
Note that this scenario is opposite to polaron destabilization in the $\omega_0=0$-limit, where a large polaron evolves into a free electron at weaker coupling $\lambda <1$ (cf. $R(\omega_0=0)$ in Fig.~\ref{Fig1}).

\begin{figure}
\includegraphics[width=\linewidth]{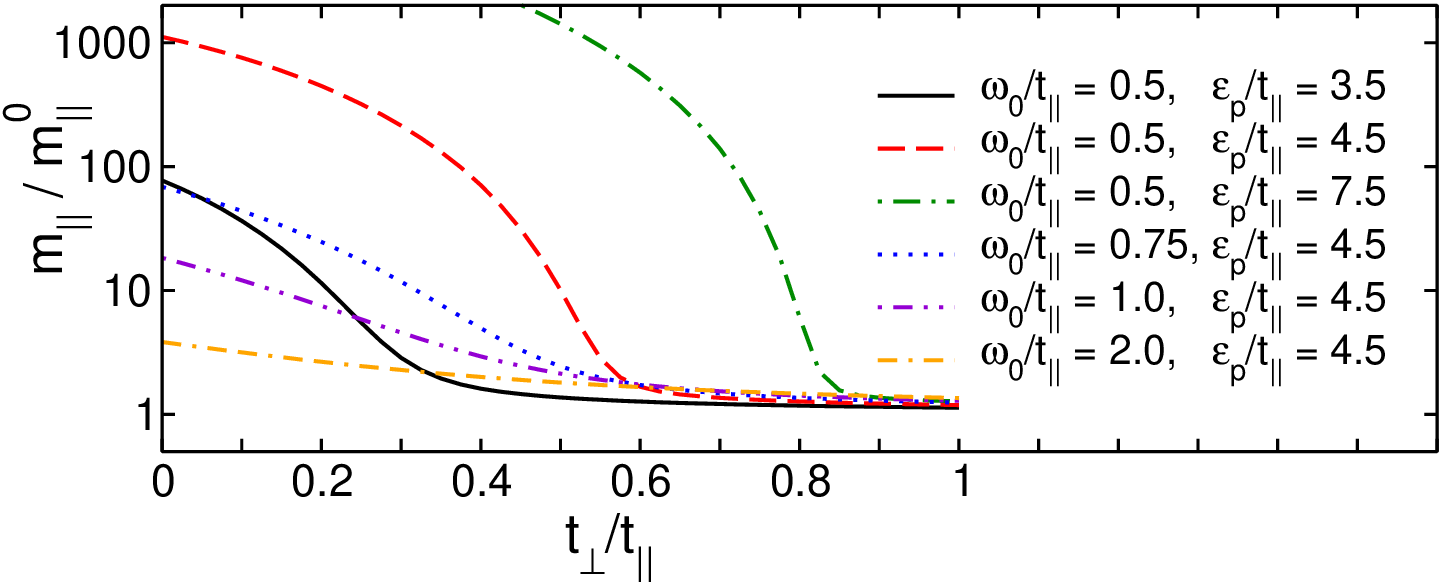}\\[0.1ex]
\caption{(Color online) Polaron mass $m_\parallel$ as a function of
  anisotropy $t_\perp/t_\parallel$,
for $\omega_0/t_\parallel$ and $\varepsilon_p/t_\parallel$ as
indicated.}
\label{Fig5}
\end{figure}

In Fig.~\ref{Fig5} we show $m_\parallel$ as a function of
$t_\perp/t_\parallel$ for fixed $\varepsilon_p$.
Generally, the polaron mass $m_\parallel$ decreases with increasing $t_\perp$.
For larger phonon frequencies, $m_\parallel$ depends only weakly on
$t_\perp/t_\parallel$, in agreement with our considerations above.
For small phonon frequency ($\omega_0/t_\parallel=0.5$), we see how a
heavy 1D polaron with $m_\parallel/m^0_\parallel \gg 1$ evolves into a light 3D particle with $m_\parallel/m^0_\parallel \approx 1$ as $t_\perp/t_\parallel$ grows from zero to one.
This behaviour occurs if the coupling is sufficiently large to create a heavy polaron in 1D (i.e. $\varepsilon_p/2 t_\parallel \gg 1$) but below the heavy polaron crossover region
$\lambda \simeq 1$ in 3D seen in the left panel in Fig.~\ref{Fig4} (i.e. $\varepsilon_p/6 t_\parallel < 1$).
Exactly then the ratio $R_\perp/R_\parallel$ interpolates between the strong- and weak-coupling curve (cf. Fig.~\ref{Fig3}).
Based on this condition a rough estimate for the $t_\perp/t_\parallel$ value when the particle finally becomes light (i.e. $m_\parallel/m^0_\parallel \approx 1$)
 is provided by the comparison of coupling energy to kinetic energy,
yielding $\varepsilon_p /2(t_\parallel+2t_\perp) \gtrsim 1$ or $t_\perp/t_\parallel \gtrsim (\varepsilon_p/t_\parallel-2)/4$,
e.g. $t_\perp/t_\parallel \gtrsim 0.5$ for $\varepsilon_p/t_\parallel=4.5$.
This estimate does essentially not depend on phonon frequency.
Especially, our findings persist also for $\omega_0/t_\parallel < 0.5$.

The observed evolution from a heavy to a light particle at small-to-intermediate phonon frequency is the equivalent of polaron destabilization in the $\omega_0=0$-limit.
We however note significant differences to the $\omega_0=0$ scenario.
First, it requires coupling strengths for which the 1D polaron is not large.
If the coupling is weak enough to allow for a large 1D polaron, no significant change of the polaron mass can occur on increasing $t_\perp/t_\parallel$ since the 1D mass is already of the order $m_\parallel/m^0_\parallel \simeq 1$.
Recall that we exclude the extreme adiabatic regime, where this statement may be violated for extremely small phonon frequencies, from our considerations.
Second, in contrast to the behaviour at zero phonon frequency, the change in $m_\parallel$ takes place in a large range
of $t_\perp/t_\parallel$, and does not occur as a rapid crossover in the vicinity of a small $t_\perp/t_\parallel$-value.
These differences clearly show that
the concept of polaron destabilization in quasi-1D systems is only of restricted relevance at finite phonon frequencies.
This is a consequence of the fact that the arguments for $\omega_0=0$
rely on the existence of large polarons.
As discussed before, those fail to exist
even at small phonon frequencies $\omega_0/t_\parallel \lesssim 0.1$.

A repeated question raised for quasi-1D systems is whether
a local or isotropic electron-phonon interaction can enhance the
anisotropy in the electronic transfer integrals, leading to
$m_\perp/m_\parallel > (t_\perp/t_\parallel)^{-1}$. 
Large anisotropies in polaron mobility that
exceed the estimates from, e.g., band structure calculations by
several orders of magnitude can be explained if acoustic phonons are
taken into account~\cite{Go84}.
It is common belief that a similar mechanism 
is absent in the Holstein model due to the purely local interaction
with isotropic optical phonons.

As a measure for the anisotropy of mass renormalization, we define
\begin{equation}\label{EqA}
  A = \frac{m_\perp/m^0_\perp - m_\parallel/m^0_\parallel
  }{m_\parallel/ m^0_\parallel} \;,
\end{equation}
or equivalently $m_\perp/m_\parallel = (1+A) m^0_\perp/m^0_\parallel$.
In the anti-adiabatic strong-coupling limit, with $m_i/m_i^0 =
\exp(g^2)$, the mass
renormalization is isotropic ($A=0$), but for smaller phonon frequency
$A\ne 0$ is possible. 
An enhancement of anisotropy corresponds to $A>0$.

\begin{figure}
\includegraphics[width=\linewidth]{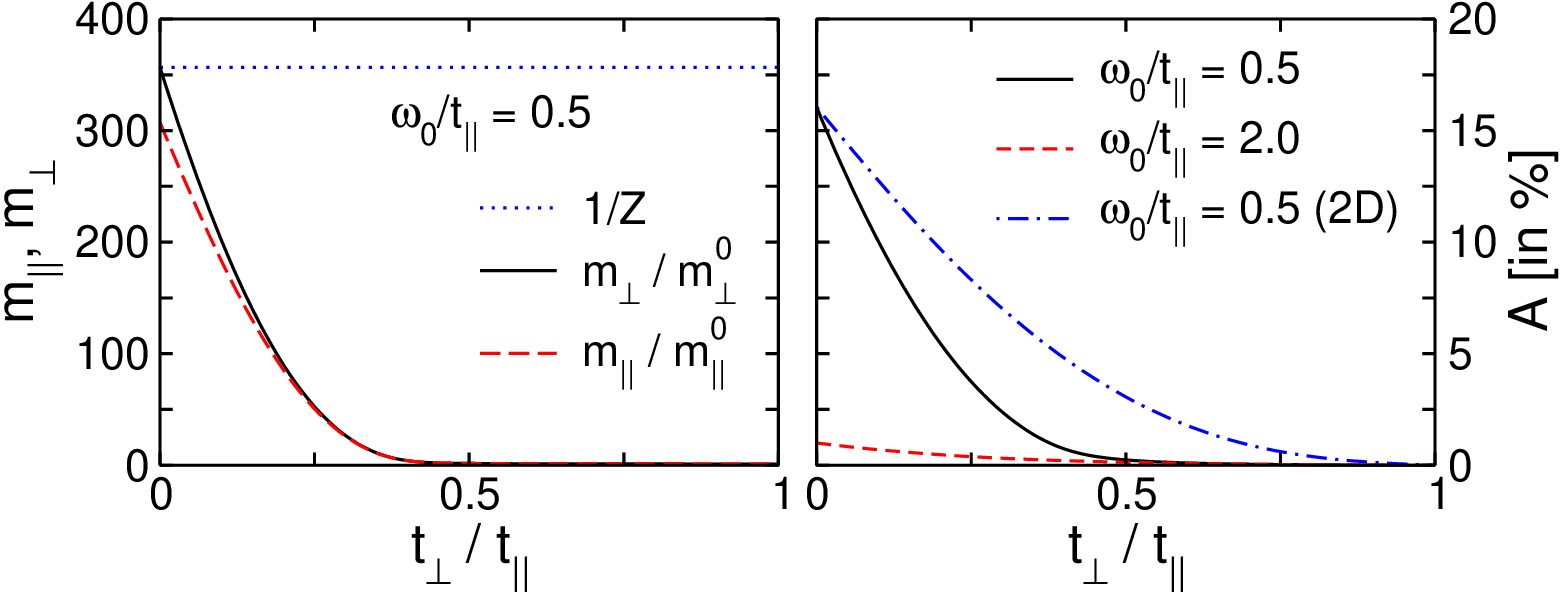}
\caption{(Color online) 
$m_\parallel$, $m_\perp$ and mass renormalization
anisotropy $A$ as a function of $t_\perp/t_\parallel$,
for $\varepsilon_p/t_\parallel=4.0$.
Left panel: For $t_\perp/t_\parallel \to 0$, the renormalization
$m_\perp/m^0_\perp$ is given by the inverse quasiparticle weight $1/Z$
(dotted line).
Right panel: In addition, $A$ is shown for a 2D system.
}
\label{Fig6}
\end{figure}

In Fig.~\ref{Fig6} we show $m_\parallel$, $m_\perp$
and the ratio $A$ for three choices of parameters.
We see that generally $m_\perp > m_\parallel$, and $A$ is maximal for
$t_\perp=0$, while of course $A=0$ for $t_\perp/t_\parallel=1$.
Evidently, $A$ is largest for small phonon frequency and large coupling, when
deviations from the anti-adiabatic limit with $A=0$ are most significant.

From first order perturbation theory in $t_\perp$ we conclude that
$A \to \delta$ for $t_\perp\to 0$,
where
\begin{equation}\label{EqDelta}
  \delta = \frac{(m^0_\parallel/m_\parallel)_{t_\perp=0} - Z}{Z}
\end{equation}
is the fractional difference between the quasiparticle weight
$Z=|\langle \psi_0| c^\dagger_{\mathbf{k}=0} |\mathrm{vac}\rangle|^2$ at
momentum $\mathbf{k}=0$ ($|\psi_0\rangle$ denotes the $\mathbf{k}=0$
groundstate) and the inverse mass renormalization
$(m^0_\parallel/m_\parallel)_{t_\perp=0}$ at $t_\perp=0$.
This relation holds independent of dimension.
As a consequence of the momentum dependence of the polaron
self-energy, $\delta \ne 0$ is possible~\cite{KTB02}.
That $\delta>0$, and thus $A>0$, results from 
effective long-range hopping processes induced by
electron-phonon-coupling, which reduce
$(m^0_\parallel/m_\parallel)_{t_\perp=0}$ compared to $1/Z$.
In a certain sense, $\delta$ measures the deviation from
the anti-adiabatic strong-coupling result $m_i/m_i^0 = Z^{-1} =
\exp(g^2)$, when $\delta=0$.

\begin{figure}
\includegraphics[width=\linewidth]{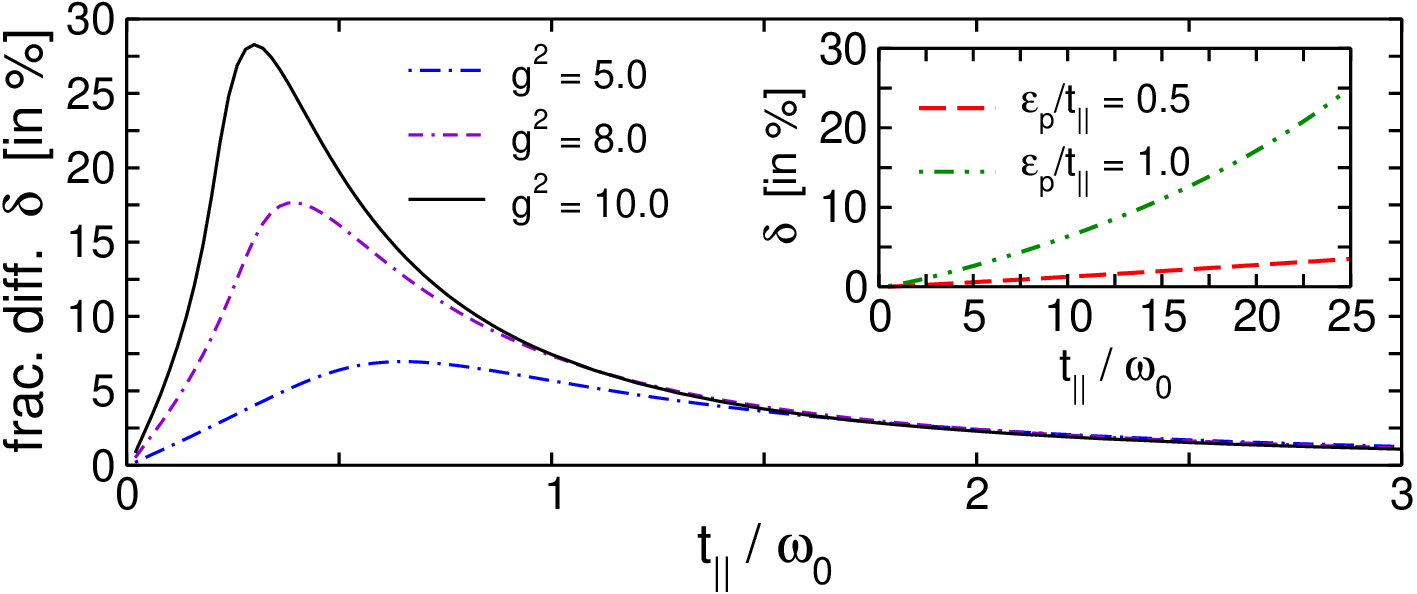}
\caption{(Color online) 
Fractional difference $\delta$ between $m_\parallel/m^0_\parallel$ and
$Z$ as a function of $\omega_0/t_\parallel$, in 1D with $t_\perp=0$, for fixed
$g^2=\varepsilon_p/\omega_0$, or fixed $\varepsilon_p/t_\parallel$ in the inset.
}
\label{Fig7}
\end{figure}

In Fig.~\ref{Fig7} we show $\delta$ for fixed
$g^2$ as a function of $\omega_0$.
For $\omega_0/t \gg 1$, the ratio $\delta$ is close to zero as the
anti-adiabatic limit is approached.
For $\omega_0 \to 0$ and fixed $g^2$, the ratio $\delta$ converges to
zero since the weak coupling regime is reached.
Decreasing $\omega_0$ or increasing $\varepsilon_p$, the ratio
$\delta$, and thus $A$, can be made very large.
For the parameters plotted, $A$ can attain values as large as $27\%$.
Note that $A$ is already large for small coupling $\varepsilon_0/t_\parallel$ provided $\omega_0/t_\parallel$ is small (see inset in Fig.~\ref{Fig7}), which implies that the mechanism described here is relevant for realistic materials.

\section{Conclusions}
To sum up, the exact results provided in this work
help to clarify the issue of polaron formation in anisotropic 3D
materials, including the 1D and isotropic 3D case.
In the extreme adiabatic regime, where large heavy 1D polarons exist for tiny phonon frequencies $\omega_0/t_\parallel \ll 0.1$,
the scenario of polaron destabilization given by Emin~\cite{Emi86} applies.
We present complementary results for the broader regime of small-to-intermediate phonon frequency and small-to-strong coupling strength.
In combination with Emin's work at least the most fundamental answers to the Holstein polaron anisotropy problem are thereby given. 
The regime of very strong coupling or large phonon frequency, where a polaron is always small or its properties do not depend on dimension, deserves no further exploration in that context.

Our findings have two major implications.
First, 
polaron destabilization occurs only in the extreme adiabatic regime, but the concept does not carry over to finite phonon frequencies 
without significant modifications.
Although it is possible to drive a heavy 1D polaron into
a light 3D particle by increasing the perpendicular electronic
transfer integral, a smooth crossover takes place instead of an instability. 
In any case, polarons in quasi-1D materials can be described by the 1D
Holstein model for small perpendicular transfer integrals.
The second implication is that even a local
Holstein interaction enhances the anisotropy in electron motion.
For small phonon frequencies this enhancement is of significant size already at weak-to-moderate coupling.
As a question for future research
we mention that the anisotropy enhancement becomes even more important
for longer-ranged electron-phonon coupling.
We stress that the frequent assumption
that the Holstein interaction leads only to isotropic changes of
material properties is not true.
Experimentally observed anisotropic behaviour provides no indication against a dominating short-range Holstein-like coupling.

\section*{Acknowledgments}
This work was supported by Deutsche Forschungsgemeinschaft through SFB 652, 
and the US Department of Energy, Center for Integrated Nanotechnologies, 
at Los Alamos National Laboratory (Contract DE-AC52-06NA25396) and 
Sandia National Laboratories (Contract DE-AC04-94AL85000).

\end{document}